\documentclass[12pt]{article}
           
\usepackage{epsfig,latexsym,amssymb,amsmath, graphics}

\def\tr{\operatorname{tr}\,}

\def\dist{{\rm dist}\,}
\def\supp{\operatorname{supp}\,}
\def\Ind{\operatorname{Ind}}
\def\e{{\rm e}}
\def\i{{\rm i}}

\def\slim{\mathop{\mathrm{s-}}\!\lim}

\newcommand{\widec}{\smash{\raisebox{18pt}{\rotatebox{180}{$\widehat{}$}}}}
\newcommand{\widecheck}[1]{\hskip 5pt\widec \hskip -5pt {#1}}

\def\etilde{\widetilde{\phantom{U}}}
\def\echeck{\widecheck{\phantom{U}}}
\def\ehat{\widehat{\phantom{U}}}
\newcommand{\tl}[2]{\genfrac{}{}{0pt}{}{#1}{#2}}
\newtheorem{thm}{Theorem}
\newtheorem{alem}{Lemma}[section]
\newtheorem{lem}[thm]{Lemma}

\title{Equality of bulk and edge Hall conductance revisited}
\author{P. Elbau, G.M. Graf\\
\normalsize\it Theoretische Physik,
ETH-H\"onggerberg, CH--8093 Z\"urich}

\begin{document}
\maketitle
\vspace{0.4cm}
\begin{abstract} 
The integral quantum Hall effect can be explained either as resulting from 
bulk or edge currents (or, as it occurs in real samples, as a combination of 
both). This leads to different definitions of Hall conductance, which agree 
under appropriate hypotheses, as shown by Schulz-Baldes et al. by means of 
$K$-theory. We propose an alternative proof based on a generalization of the 
index 
of a pair of projections to more general operators. The equality of 
conductances is an expression of the stability of that index as a flux tube 
is moved from within the bulk across the boundary of a sample. 
\end{abstract}
\bigskip

\section*{The model and the result}

The simultaneous quantization of bulk and edge conductance is essential to the QHE in finite samples, as explained in \cite{F,KRSa}. In these two references that property is established in the context of an effective field theory description, resp. of a microscopic treatment suitable to the integral QHE. 
The present paper is placed in the latter setting as well.\\

In our model $H$ is a discrete Schr\"odinger operator on the single-particle Hilbert space $\ell^2({\mathbb Z} \times {\mathbb N})$ over the upper half-plane. It is obtained
from the restriction (with e.g. Dirichlet boundary conditions) of a
`bulk' Hamiltonian $H_B$ acting on 
$\ell^2({\mathbb Z} \times {\mathbb Z})$. These assumptions are spelled out 
in detail at the end of this
section. The spectrum of $H_B$ (but not that of $H$, as a rule) has an open
gap $\Delta$ containing the Fermi energy: 
\begin{equation}
\Delta \cap \sigma (H_B)=\varnothing\;.
\label{eq:0} 
\end{equation}
Let $P_B$ be the Fermi projection:
$P_B=E_{(-\infty,\mu]}(H_B)$ for any $\mu \in \Delta$.

A real-valued function $g\in C^\infty ({\mathbb R})$ with
$g(\lambda)= 1$ (resp. 0) for $\lambda$ large and negative (resp. positive)
will be called a {\it switch function}.
We remark that $P_B = g(H_B)$ if the switch function has 
$\supp g' \subset\Delta$.

\begin{thm}\label{thm} Assume the hypotheses as described and, in particular, (\ref{eq:0}). Let
\begin{equation}
\sigma_B=\frac{1}{2\pi}\;\Ind(UP_BU^*,P_B)\;,
\label{eq:1a}
\end{equation}
where $U=U (\vec{r}) = \e^{\i\arg{\vec{r}}}$ be the {\bf
bulk Hall conductance}; and let
\begin{equation}
\sigma_E=- \tr ( g'(H)\i[H,\chi (x)])\;,
\label{eq:1}
\end{equation}
where $g$ and $\chi$ are switch functions with 
$\supp g' \subset \Delta$, be the {\bf edge Hall conductance}. Then
\begin{equation}
\sigma_B=\sigma_E\;.
\label{eq:2}
\end{equation}
In particular, $\sigma_E$ is independent of $g$ and $\chi$ as stated.
\end{thm}

\noindent
{\bf Remarks.} 1) $\Ind(P,Q)$ is the index of a pair of projections, 
see \cite{AS2a}, from where also the definition of $\sigma_B$ is taken, except
for a change of sign. In other words, their definition of $\sigma_B$ agrees 
with the Kubo formula (6.18) for $\sigma_{12}$, whereas ours with 
$\sigma_{21}$. Or equivalently: their definition is such that for a Landau 
Hamiltonian with 
magnetic field $B >0$ and electron charge $e=+1$ one has $\sigma_B >0$, see 
Remark 6.7c. Ours is opposite.

2) $U(\vec{r})$ can be replaced, without affecting $\sigma_B$, by
\begin{equation}
U(\vec{r}) = \e^{\i\varphi (\arg{\vec{r}})}\;,
\label{eq:3}
\end{equation}
where $\varphi: S^1\to S^1$ is a continuous function with winding number 1. 
This follows by continuity from the additivity \cite{AS2a} and stability of the index:
\[
\Vert Q-P\Vert < 1\; \Rightarrow\; \Ind(Q,P)=0\;.
\]

3) The rationale for the definition (\ref{eq:1}) is that $-\i[H,\chi (x)]$ is
the current operator in $x$-direction (for $\chi (x)=\theta(-x)$, it is the
current across $x=0$). For $-g' (H)=E_{[\mu_1,\mu_2]}(H)/(\mu_2-\mu_1)$
(\ref{eq:1}) is (up to the sign) the expected current in 1-particle density 
matrix $E_{[\mu_1,\mu_2]}(H)$, corrisponding to filled edge levels 
$[\mu_1,\mu_2]\subset\Delta$, divided by the potential difference. 
For the above Landau Hamiltonian the current is positive, since the electrons
run in the positive $x$-direction near the boundary. Thus $\sigma_E$ is, 
like $\sigma_B$, negative.\\

The result (\ref{eq:2}) was proven in \cite{KRSa} and, more extensively, in
\cite{KRSb} using non-commutative
geometry and $K$-theory. (However, the quantization of $\sigma_E$ was shown there without making use of these tecniques). The present proof makes use of basic
functional analysis. While their result is established using and extending  tools developed in \cite{Bel}, ours bears a similar relation to \cite{AS2a}.\\

We conclude this section by specifying the Schr\"odinger Hamiltonians
$H$ used here. Lattice points are denoted as $\vec{r}=(x,y)$, 
corresponding Kronecker states as 
$\delta_{\vec{r}}\in\ell^2({\mathbb Z} \times {\mathbb N})$ and matrix 
elements as $H(\vec{r}_1,\vec{r}_2)=(\delta_{\vec{r}_1},H\delta_{\vec{r}_2})$.
We assume $H$ to be a self-adjoint operator with short-range off-diagonal
hopping terms:
\begin{equation}
\sup_{\vec{r}_1}\sum_{\vec{r}_2}|H(\vec{r}_1,\vec{r}_2)|
(\e^{\mu_0|\vec{r}_1-\vec{r}_2|}-1)<\infty
\label{eq:srh}
\end{equation}
for some $\mu_0>0$. The bulk Hamiltonian $H_B$ is of the same form, except 
that the lattice is
${\mathbb Z} \times {\mathbb Z}={\mathbb Z}^2$. It should restrict to $H$ on 
the upper half-plane under some largely arbitrary boundary condition. More 
precisely, let
$J: \ell^2({\mathbb Z} \times {\mathbb N})\to\ell^2({\mathbb Z}^2)$ 
denote the extension by $0$. We assume that the `edge term'
\[
E=J H-H_BJ:\; \ell^2({\mathbb Z} \times {\mathbb N})\to
\ell^2({\mathbb Z}^2 )
\]
satisfies 
\begin{equation}
\sum_{\vec{r}^{\,\prime}\in{\mathbb Z}\times{\mathbb N}}
|E(\vec{r},\vec{r}^{\,\prime})|\le C\e^{-\mu_0|y|}
\label{eq:Eest}
\end{equation}
for all $\vec{r}=(x,y)\in{\mathbb Z}^2$. For instance, for Dirichlet boundary 
conditions,
\[
E(\vec{r},\vec{r}^{\,\prime})=\begin{cases} 
-H_B(\vec{r},\vec{r}^{\,\prime})\;,\qquad&(y<0)\;,\\
0\;,\qquad&(y\ge 0)\;,
\end{cases}
\]
whence (\ref{eq:Eest}) follows from (\ref{eq:srh}) for $H_B$ at the expense of
making $\mu_0$ smaller.\\

The trace ideals
of operators on the Hilbert space $\ell^2({\mathbb Z} \times {\mathbb N})$ or 
$\ell^2({\mathbb Z}^2)$, depending on the context, are 
denoted as ${\mathcal J}_p$, ($1\le
p<\infty$), with norm $\|\cdot\|_p$. Universal constants are denoted by $C$.

\section*{Idea and outline of the proof}

We consider the gauge transformation (\ref{eq:3}) with $\varphi$ having 
$\supp\varphi' \subset [\pi/4, 3\pi/4]$, so that $U(\vec{r})-1$ is
supported in a wedge pointing upwards. We shall compare two modifications 
thereof. The first one, $\widecheck{U}_a$, is obtained from (\ref{eq:3}) by 
changing $U(\vec{r})$ to 1 for $y <a$. 

\begin{figure}[h]
\begin{center}

\input{befig.pstex_t}
 
\caption{Isolines of $U,\,\protect\widecheck{U}_a,\,\widehat{U}_a$}
\label{figure:fig1}
\end{center}
\end{figure}

The second one, $\widehat{U}_a$, is obtained from $\widecheck{U}_a$ by pulling
the line of fluxes at $y=a$ across the boundary, as in the figure. \\

Morally, the Hall conductance $\sigma_B$ is given as
\begin{equation}
\frac{1}{2\pi} \tr( \widetilde{U}_a g(H) \widetilde{U}_a^*-g(H))
\label{eq:4}
\end{equation}
with either $\etilde = \echeck,\,\ehat$. Indeed, in both cases the 
heuristic argument, explained in more detail in \cite{AS2a}, Sect.~5, is that 
the trace in (\ref{eq:4}) counts the number of electrons which are pulled to
infinity as the gauge field is switched on
adiabatically starting from zero to a flux quantum, see 
Fig.~\ref{figure:fig1}. That number may also by 
computed by integrating the current 
\begin{equation}
\vec{\jmath}=\sigma_B\varepsilon\vec{E}
\label{eq:hall}
\end{equation}
(with $\varepsilon$ denoting a rotation by $\pi/2$) over time and across a 
large circle $\mathcal C$ enclosing the flux. Here 
$\vec{E}=\i\partial_t\nabla(\log\widetilde{U}_a)$ is the electric field 
accompanying the change of magnetic
field, and is the same on $\mathcal C$ in the two cases. Since the 
phenomenological equation (\ref{eq:hall}) is valid only well inside the sample, it is crucial that the isolines of the gauge transformation run to infinity 
through the upper half-plane, so that $\vec{E}$ vanishes where $\mathcal C$
crosses the boundary of the sample. 

It appears reasonable, even without recourse to (\ref{eq:hall}), that 
for $\etilde = \echeck$ and $a\to\infty$ (\ref{eq:4}) tends to $\sigma_B$ as 
defined in (\ref{eq:1a}). As for $\widehat{U}_a$ note that
\[
\widehat{U}_a (\vec{r}) = \e^{2\pi\i \chi_a (\vec{r})}, 
\]
where $\chi_a(\vec{r})$ is a single-valued function over the sample and, for 
$\vec{r}$ close to the boundary,
\[
\chi_a (\vec{r}) =\chi_a(x) =\chi (x/a)
\]
is a switch function. This suggests that
\begin{eqnarray*}
\frac{1}{2\pi} \tr\bigl( \widehat{U}_a g (H) \widehat{U}_a^* - g(H)\bigr) 
&=& \frac{1}{2\pi} \int_0^{2\pi} d \varphi \frac{d}{d\varphi} 
\tr\bigl[ \e^{\i\varphi \chi_a} g (H) \e^{-\i\varphi\chi_a} - g (H) \bigr] \\
&=& - \tr\bigl( \i [ g(H), \chi_a]\bigr)
= - \tr(g'(H) \i [H, \chi_a]) =\sigma_E\;,
\end{eqnarray*}
where the last two traces are formally equal since the operators inside 
differ by a commutator.

The trouble with this explanation for $\sigma_B= \sigma_E$
is that none of the traces starting with
(\ref{eq:4}), except for the last one, is well-defined. In fact, one
has the weaker property $\widetilde{U}_a g(H) \widetilde{U}_a^* - g (H) \in
{\mathcal J}_3$ for $g$ a switch function (but notice that as a rule
even this fails if $g$ is taken as a step function, a fact related to
Theorem~3.11 in \cite{AS2a}). 

Put differently: the formal eigenvalue sum represented by (\ref{eq:4})
is not absolutely convergent, but exhibits strong cancellations
between small eigenvalues of opposite sign (which are exact except for
$\lambda = \pm 1$ in a bulk situation, where $g(H_B)=P_B$ is a
projection \cite{AS2b}). Let therefore $f_t(\lambda)$ be an odd
function with $f_t(1)=1$ interpolating between $\lambda^3$ (as $t=0)$
and $\lambda$ (as $t=\infty$). For definiteness we take
\begin{equation}
f_t(\lambda)=\frac{(1+t) \lambda^3}{1+t \lambda^2}\;.
\label{eq:5}
\end{equation}
We regard $\lim_{t\to\infty} \tr f_t(A)$ as a replacement for
$\tr A$, when the latter is not defined. But first we pass to a more
general setting. \\

We consider a fixed bounded operator $P$ (typically not a projection!) on a 
Hilbert space 
${\mathcal H}$ equipped with a fixed orthonormal basis ${\mathcal B}$. Our
standing assumptions are: let
\begin{equation}
Q=U P U^*\;,
\label{eq:6o}
\end{equation}
where $U$ is a unitary operator satisfying
\begin{eqnarray}
&{\mathcal B}\hbox{ is an eigenbasis for }U\;,&
\label{eq:6a}\\
&Q-P \in {\mathcal J}_3\;,&
\label{eq:6b}\\
&(Q-P) (P-P^2)\;,\; (P-P^2) (Q-P) \in {\mathcal J}_1\;,&
\label{eq:6c}\\
&p(Q) - p(P) \in {\mathcal J}_1\;,&
\label{eq:6d}
\end{eqnarray}
for any polynomial $p(\lambda)$ with $p(0)=p(1)=0$ and $\deg p\le 3$. This 
implies
\begin{equation}
\tr \bigl( p (Q) - p (P)\bigr) =0\;, 
\label{eq:38}
\end{equation}
as it is seen be evaluating the trace in an eigenbasis of
$U$. Specifically, (\ref{eq:38}) will be used for the polynomials 
$p (\lambda) = \lambda - \lambda^2$ and 
$p (\lambda)=(1-2\lambda)(\lambda-\lambda^2)=
\lambda-3\lambda^2+2\lambda^3$,
which span the above space of polynomials.\\

As an abstract replacement for (\ref{eq:4}) we have 

\begin{lem}\label{lm1} Assume (\ref{eq:6o}--\ref{eq:6d}) and $P=P^*$. Then 
\begin{equation}
\lim_{t\to\infty}\tr f_t (Q-P) 
= \tr \Bigl( \frac 3 2 \bigl\{ Q-P, (Q-Q^2) + (P-P^2)\bigr\} +
(Q-P)^3\Bigr) \equiv K (U)\;.
\label{eq:39}
\end{equation}
\end{lem}

The proof of Theorem~\ref{thm} will not depend on Lemma~\ref{lm1}, except for 
the fact that $K(U)$ is well-defined. The limit (\ref{eq:39}) will thus be
proved only towards the end of the paper.\\

The heuristic discussion following (\ref{eq:4}) is now substantiated in terms
of $K(U)$.

\begin{lem}\label{lm2} Let $Q_i=U_iPU_i^*$, $(i=1,2)$ satisfy
(\ref{eq:6a}--\ref{eq:6d}) and assume
\begin{equation}
U_2-U_1 \in {\mathcal J}_1\;.
\label{eq:40}
\end{equation}
Then $K(U_1)=K(U_2)$.
\end{lem}

We now turn to the application to the quantum Hall effect.

\begin{lem}\label{lm3} i) The assumptions (\ref{eq:6a}--\ref{eq:6d}) hold 
true for ${\mathcal H} = \ell^2 ({\mathbb Z} \times {\mathbb N}), B =\{
\delta_{\vec{r}}\}_{\vec{r}\in {\mathbb Z} \times {\mathbb N}}$,
\begin{equation}
P = g (H)\;,\qquad U = \widetilde{U}_a\;,\qquad 
Q_a = \widetilde{U}_ag (H) \widetilde{U}_a^*
\label{eq:21}
\end{equation}
with $\etilde = \echeck,\,\ehat$ and $g$ as in Theorem~\ref{thm}.

ii) Assumption (\ref{eq:40}) applies to $U_i=\widetilde{U}_a$, with
separate choices of $a$ and $\etilde = \echeck,\,\ehat$ for
$i=1,2$. 
\end{lem}

Therefore, $K(\widetilde{U}_a)$ is independent of $a$ and $\etilde$. 

\begin{lem}\label{lm4} Let (\ref{eq:21}) with $\etilde = \echeck$. Then
\begin{eqnarray}
&&\lim_{a\to\infty} \tr (Q_a-P)^3= 2\pi\sigma_B\;,
\label{eq:41}\\
&&\lim_{a\to\infty} \frac 3 2 \,\tr \{ Q_a-P, (P-P^2) +
(Q_a-Q_a^2)\}= 0\;.
\label{eq:42}
\end{eqnarray}
\end{lem}
\begin{lem}\label{lm5} Let (\ref{eq:21}) with $\etilde = \ehat$. Then
\begin{eqnarray}
&&\lim_{a\to\infty} \tr (Q_a-P)^3=0 \;,
\label{eq:43}\\
&&\lim_{a\to\infty} \frac 3 2 \,\tr \{ Q_a-P, (P-P^2) +
(Q_a-Q_a^2)\}=2\pi\sigma_E \;.
\label{eq:22}
\end{eqnarray}
\end{lem}

\noindent
{\bf Proof} of Theorem~\ref{thm}. Is immediate from 
Lemmas~\ref{lm2}--\ref{lm5}.\hfill$\square$ 

\section*{The details}

The starting point to the proofs of Lemma~\ref{lm1} and \ref{lm2} are two 
identities from \cite{AS2a} valid for projections $P=P^2$ and $Q=Q^2$. They
are
\begin{eqnarray}
&(Q-P)-(Q-P)^3=[QP,PQ]=[QP,[P,Q-P]]\;,&
\label{eq:10}\\
&[P,(Q-P)^2]=[Q,(Q-P)^2]=0\;.&
\label{eq:11}
\end{eqnarray}
The first was used there for it yields the case $n=0$ of
\[
\tr (Q-P)^{2n+3}=\tr (Q-P)^{2n+1}
\]
for $Q-P\in{\mathcal J}_{2n+1}$. The second yields the extension to
$n\in{\mathbb N}$. For later purpose we remark that they similarly
yield
\begin{equation}
\tr f_t (Q-P)=\tr (Q-P)^3
\label{eq:12}
\end{equation}
for $0\le t <\infty$ if $P-Q\in {\mathcal J}_3$. Indeed: since
\begin{equation}
f_t (\lambda) -\lambda^3= \frac{t\lambda^2}{1+t\lambda^2}
(\lambda-\lambda^3) 
\label{eq:13}
\end{equation}
we have
\[
f_t (Q-P)-(Q-P)^3=t\bigl[QP, [P, \frac{(Q-P)^3}{1+t(Q-P)^2}]\bigr]
\]
with the inner commutator being trace class, whence (\ref{eq:12}). 

Our primary concern here is however a generalization of (\ref{eq:10},
\ref{eq:11}) to arbitrary bounded operators $P,Q$. More precisely, we
take the half-difference between (\ref{eq:10}) and its
``particle-hole'' reversed variant ($P\to1-P,\,Q\to1-Q$), and correct
the result by the appropriate terms involving $P-P^2$ and $Q-Q^2$:
\begin{eqnarray}
(Q-P)-(Q-P)^3 
&=& \frac 1 2 [QP,PQ]-\frac 12 \bigl[
(1-Q)(1-P),(1-P)(1-Q)\bigr]\nonumber\\
&&+ (1-2Q)(Q-Q^2)-(1-2P)(P-P^2)\nonumber\\
&&+\frac 32 \{ Q-P, Q-Q^2 + P-P^2\}\;.
\label{eq:14}
\end{eqnarray}
In the new setting (\ref{eq:11}) is replaced with 
\begin{equation}
\bigl[ P, (Q-P)^2\bigr]=\bigl[ Q,(Q-P)^2\bigr]=\bigl[
Q-P,(Q-Q^2)-(P-P^2)\bigr]\;. 
\label{eq:15}
\end{equation}
These relations are conveniently stated in terms of the operators
\begin{equation}
A=Q-P\;,\qquad B=1-P-Q
\label{eq:16}
\end{equation}
introduced in \cite{K,AS2b}, for which
\begin{eqnarray}
\{A,B\} &=& 2\bigl[ (Q-Q^2)-(P-P^2)\bigr]\;,
\label{eq:17}\\
1-A^2-B^2&=&2\bigl[ (Q-Q^2)+(P-P^2)\bigr]\;.
\label{eq:18}
\end{eqnarray}
Then (\ref{eq:14}) reads (with equality line by line)
\begin{eqnarray}
A-A^3&=& \frac 14\bigl[B,[B,A]\bigr]\nonumber\\
&&+ \frac 14\bigl\{B, \{ A,B\}\bigr\} 
- \frac 1 4 \bigl\{ A,1-A^2-B^2\bigr\} \nonumber\\
&&+ \frac 34 \bigl\{ A, 1-A^2-B^2\bigr\}
\label{eq:19}
\end{eqnarray}
and (\ref{eq:15}) (after multiplication by 2)
\begin{equation}
[A^2,B]=[A,\{A,B\}]\;.
\label{eq:20}
\end{equation}

\noindent
{\bf Proof} of Lemma~\ref{lm2}. We remark that
\[
\bigl\{ Q-P,(Q-Q^2)+(P-P^2)\bigr\} = 2\bigl\{ Q-P,P-P^2\}+\{Q-P,
p(Q)-p(P) \bigr\}
\]
with $p(\lambda)=\lambda-\lambda^2$, is trace class by our assumptions
(\ref{eq:6c}, \ref{eq:6d}). Thus $K(U)$ in (\ref{eq:39}) is well-defined. Let
$A_i=Q_i-P$, $(i=1,2)$, and similarly for $B_i$. We take the
difference between (\ref{eq:19}) (or (\ref{eq:14})) in the two
cases. In a mixed notation we have
\begin{eqnarray}
A_i\bigm|_1^2 - A_i^3\bigm|_1^2 
&=& \frac 14 \bigl[ B_i,[B_i,A_i]\bigr]\bigm|_1^2\nonumber\\
&&+ \bigl[ p(Q_i)-p(P)\bigr]\bigm|_1^2\nonumber\\
&&+ \frac 32 \bigl\{ Q_i-P,(Q_i-Q_i^2)+(P-P^2)\bigr\}\bigm|_1^2
\label{eq:44}
\end{eqnarray}
with $p(\lambda)=(1-2\lambda)(\lambda-\lambda^2)$. We note that
$A_2-A_1 =-(B_2-B_1) = Q_2-Q_1 \in {\mathcal J}_1$ with $\tr
(Q_2-Q_1)=0.$ Indeed, by (\ref{eq:40}), 
\begin{equation}
Q_2-Q_1 = U_2PU_2^*-U_1PU_1^*
= (U_2-U_1) PU_2^* + U_1P(U_2-U_1)^*
\label{eq:45}
\end{equation}
is trace class, and the trace is seen to vanish using the basis
${\mathcal B}$. Writing
\[
\bigl[ B_i,[B_i,A_i]\bigr]\bigm|_1^2= \bigl[
B_2,[B_2,A_2-A_1]\bigr] + \bigl[B_2,[B_2-B_1,A_1]\bigr] + \bigl[
B_2-B_1,[B_1,A_1]\bigr] 
\]
we see that the first term on the r.h.s. of (\ref{eq:44}) is trace
class with vanishing trace. So is the next one due to
(\ref{eq:38}). \hfill $\square$\\

\noindent
{\bf Proof} of Lemma~\ref{lm3}. Eq. (\ref{eq:6a}) is evident, since the
$\widetilde{U}_a$ are multiplication operators. Let $U(\vec{r})$ be given
by (\ref{eq:3}) as in Figure~\ref{figure:fig1}. Since $U-\widetilde{U}_a$ has 
compact support as a function, it is trace class as an
operator. Thus (ii) holds true and it suffices to prove
(\ref{eq:6b}--\ref{eq:6d}) for $U$ instead of $\widetilde{U}_a$, cf.
(\ref{eq:45}). The $(\vec{r}_1,\vec{r}_2)$-matrix element of
$(Q-P)U=Ug(H)-g(H)U$ is
\[
g(H)(\vec{r}_1,\vec{r}_2)\bigl(U(\vec{r}_1)-U(\vec{r}_2)\bigr)\;,
\]
so (\ref{eq:6b}) follows from (\ref{eq:A.4}),
\[
\big| U(\vec{r}_1)-U(\vec{r}_2)\big| \le C
\frac{|\vec{r}_1-\vec{r}_2|}{1+|\vec{r}_1|} 
\]
and (\ref{eq:A.7}) with $p=3$. To prove (\ref{eq:6d}), we note that
$G=p\circ g$ has $\supp G \subset \Delta$. Hence (\ref{eq:A.6})
applies. Writing the matrix element of $(p(Q)-p(P)) U=U G(H)-G(H)U$ as
before, the claim follows. As mentioned, the verification of
(\ref{eq:6c}) could equally be done on the basis of $U$ instead of
$\widetilde{U}_a$. However we prefer to do this for $\etilde =\echeck,\ehat$
explicitly, since this will provide estimates, stated in the lemma
below, which will be useful in the proofs of Lemma~\ref{lm4},~\ref{lm5}. 
Technically, the first part of (\ref{eq:6c}) is just the case $b=0$ in
(\ref{eq:27}) below. The second part follows by taking the adjoint.
\hfill $\square$\\

The rough reason for
\[
(Q_a-P)(P-P^2)=\bigl( \widetilde{U}_a g(H) \widetilde{U}_a^*-g(H)\bigr)
\bigl( g(H)-g(H)^2\bigr) 
\]
to be trace class is that $\supp (\widetilde{U}_a-1)$ has compact
intersection (possibly empty) with the boundary.

\begin{lem}\label{lm6} Let $F_b=F_b(y)$ be the characteristic function of the
neighborhood $\{ \vec{r}|y< b\}$ of the boundary. Then, in the
notation (\ref{eq:21}), 
\begin{eqnarray}
&&\Vert (Q_a-P)(1-F_b)(P-P^2)\Vert_1 \le C (1+a) \e^{-\kappa b}\;,
\label{eq:27}\\
&&\Vert (Q_a-P)(1-F_b)(Q_a-Q_a^2)\Vert_1 \le C(1+a) \e^{-\kappa b}
\label{eq:27a}
\end{eqnarray}
for both $\etilde=\echeck,\ehat$ and some $\kappa>0$. For $b\le a/2$ we 
furthermore have
\begin{equation}
\Vert (Q_a-P) F_b\Vert_1 \le C_N (1+a)^{-N}
\label{eq:46}
\end{equation}
in case $\etilde =\echeck$; and
\begin{eqnarray}
&&\Vert (Q_a-P) F_b\Vert_1 \le C\cdot  b\;,
\label{eq:25}\\
&&\Vert (Q_a-P) F_b\Vert_2 \le C(b/a)^{1/2}
\label{eq:26}
\end{eqnarray}
in case $\etilde =\ehat$.
\end{lem}

\noindent
{\bf Proof.} We set $G(H)=P-P^2=g(H)-g(H)^2$ and estimate
(\ref{eq:27}) as
\begin{eqnarray*}
\Vert (Q_a-P)(1-F_b) G(H)\Vert_1&=& \Vert (Q_a-P)(1-F_b) \e^{-\kappa
  y} \e^{\kappa y} G(H)\Vert_1 \\
& \le& \e^{-\frac{\kappa}{2} b} \Vert (Q_a-P)
\e^{-\frac{\kappa}{2}y}\Vert_1 \Vert \e^{\kappa y} G (H)\Vert\;,
\end{eqnarray*}
where the last norm is finite due to (\ref{eq:A.6}). The operator
$T=(Q_a-P) \e^{-\frac \kappa 2 y}\widetilde{U}_a$ has kernel
\[
T(\vec{r}_1,\vec{r}_2)=
g(H)(\vec{r}_1,\vec{r}_2) \bigl(\widetilde{U}_a (\vec{r}_1) - \widetilde{U}_a
(\vec{r}_2)\bigr) \e^{-\frac\kappa 2 y_2}
\]
with 
\begin{equation}
|\widetilde{U}_a(\vec{r}_1)-\widetilde{U}_a(\vec{r}_2)| \le C_k \bigl(
1+(|x_2|-a-y_2)_+\bigr)^{-k} \bigl( 1+ |\vec{r}_1-\vec{r}_2|\bigr)^k\;.
\label{eq:47}
\end{equation}
In fact if $|x_2|<a+y_2$, the first factor on the r.h.s. is bounded
below by $1$, while the l.h.s. is bounded above by 2. In the opposite case
$|x_2| \ge a+y_2$, we distinguish between $|x_1|\ge a+y_1$, whence
the l.h.s. vanishes (see Fig.~\ref{figure:fig1}), and $|x_1| < a+y_1$, 
where $\sqrt{2}|\vec{r}_1-\vec{r}_2| \ge |x_2|-a-y_2$ implies that the r.h.s. is bounded below away from $0$. We claim this proves
\begin{equation}
\Vert T\Vert_1 = \Vert (Q_a-P) \e^{-\frac\kappa 2 y} \widetilde{U}_a
\Vert_1 \le C(1+a)\;, 
\label{eq:48}
\end{equation}
and hence (\ref{eq:27}). To this end we apply (\ref{eq:A.7}) with
$p=1$: using (\ref{eq:A.4}) with $N+k$ instead of $N$ we have
\begin{eqnarray*}
\sum_{\vec{r}\in {\mathbb Z}\times {\mathbb N}}
|T(\vec{r}+\vec{s},\vec{r})| &
\le& C (1+|\vec{s}|)^{-N} \sum_{\vec{r}} \bigl(
1+(|x|-a-y)_+\bigr)^{-k} \cdot \e^{-\frac\kappa 2 y} \\
&\le& C (1+|\vec{s}|)^{-N} \sum_{y=0}^\infty (1+a+y) \e^{-\frac \kappa
  2 y}\le C (1+|\vec{s}|)^{-N} \cdot (1+a)\;, 
\end{eqnarray*}
for $k\ge 2$. This is summable w.r.t. $\vec{s}\in{\mathbb Z}^2$ for
$N\ge 3$. Taking (\ref{eq:27}) with $\widetilde{U}_a^*$ instead of
$\widetilde{U}_a$ yields (\ref{eq:27a}).\par
Let now $b\le a/2$ for the rest of the proof. The proof of
(\ref{eq:46}) is just like that of (\ref{eq:48}), which we supplement
with $\widecheck{U}_a (\vec{r_1})-\widecheck{U}_a(\vec{r}_2)=0$ if $y_2<a/2$ 
and $|\vec{r}_1-\vec{r}_2|\le a/2$. This yields for $T=(Q_a-P) F_b
\widecheck{U}_a$
\begin{eqnarray*}
\sum_{\vec{r}} |T(\vec{r}+\vec{s},\vec{r})|
&\le &C (1+|\vec{s}|)^{-N} \sum_{y=0}^\infty (1+a+y) F_b (y)\\
& \le &C(1+|\vec{s}|)^{-N} (1+a)^2\;,
\end{eqnarray*}
and $=0$ if $|\vec{s}|<a/2$. Thus
\[
\Vert T\Vert_1 \le C(1+a)^2 \sum_{\vec{s}:|\vec{s}|\ge a/2}
(1+|\vec{s}|)^{-N} \le C(1+a)^2 (1+a)^{-(N-2)}. 
\]
Let finally $\etilde=\ehat$, where
\begin{equation}
\big| \widehat{U}_a (\vec{r}_1) - \widehat{U}_a (\vec{r}_2)\big| \le C
\frac{|\vec{r}_1-\vec{r}_2|}{a+|\vec{r}_2|}\;.
\label{eq:49}
\end{equation}
This holds true for $a=1$ and $\vec{r}_1, \vec{r}_2 \in {\mathbb
  R}^2$, and follows by scaling, $\widehat{U}_a(\vec{r})=\widehat{U}_1
(\vec{r}/a)$, for $a>0$. To estimate $T=(Q_a-P) F_b\widehat{U}_a$ we use
(\ref{eq:49}) for $|x_2|<3a$ and (\ref{eq:47}) for $|x_2|\ge 3a$
(with $N+1$, resp. $N+k$ in (\ref{eq:A.4})). Thus
\begin{eqnarray*}
\sum_{\vec{r} \in{\mathbb Z}\times {\mathbb N}} |
T(\vec{r}+\vec{s},\vec{r})|
&\le& C (1+|\vec{s}|)^{-N} \sum_{y=0}^{b-1} \biggl( \sum_{|x|<3a}
\frac 1 a + \sum_{|x|\ge 3a} \bigl( 1+
(|x|-a-y)_+\bigr)^{-k}\biggr)\\
&\le& C(1+|\vec{s}|)^{-N} b \biggl( 6+2 \sum_{m=a}^\infty
(1+m)^{-k}\biggr)\;, 
\end{eqnarray*}
where we used $|x|-a-y\ge 3a-2a=a$. Similarly,
\begin{eqnarray*}
\sum_{r\in{\mathbb Z}\times{\mathbb N}}
|T(\vec{r}+\vec{s},\vec{r})|^2
&\le& C(1+|\vec{s}|)^{-2N} b \biggl( \frac 1 a +
\sum_{m=a}^\infty (1+m)^{-2k}\biggr) \\
&\le& C(1+|\vec{s}|)^{-2N}\cdot b/a\;.
\end{eqnarray*}
\rightline{$\square$}\\

\noindent
{\bf Proof} of Lemma~\ref{lm4}. Let 
$A=Q_a-P=\widecheck{U}_a g(H) \widecheck{U}_a^* - g(H)$,
$A_B=\widecheck{U}_a g(H_B) \widecheck{U}_a^*-g(H_B)=$ 
$\phantom{}\widecheck{U}_a P_B \widecheck{U}_a^*-P_B$ and
$D=(A-A_B) \widecheck{U}_a$, where $g(H)\equiv Jg(H)J^*$ is now meant as an 
operator on $\ell^2
({\mathbb Z}\times {\mathbb Z})$, simply extended by zero. The kernel
of $D$,
\[
D(\vec{r}_1,\vec{r}_2)=\bigl( Jg(H)J^*-g(H_B)\bigr)
(\vec{r}_1,\vec{r}_2) \bigl( \widecheck{U}_a(\vec{r}_1) - 
\widecheck{U}_a (\vec{r}_2)\bigr)\;,
\]
satisfies (up to a factor 2) the bound (\ref{eq:A.5}), and vanishes if
both $\vec{r}_1,\vec{r}_2$ are outside of the wedge. Thus
(\ref{eq:A.7}) with $p=1$ shows 
\[
\Vert D\Vert_1 \le C \e^{-\kappa a}\;.
\]
Writing $A^3-A_B^3=A^2(A-A_B)+A(A-A_B)A_B + (A-A_B)A_B^2$, this proves
\[
\lim_{a\to\infty} (\tr A^3- \tr A_B^3)=0\;.
\]
But, see \cite{AS2a}, 
\begin{equation}
\tr A_B^3=\Ind\bigl( \widecheck{U}_a P_B\widecheck{U}_a^*,P_B\bigr)
\label{eq:50}
\end{equation}
is independent of $a$ due to the stability of the index (\cite{K},
Theorem~5.26) under compact perturbations (or use Lemma~\ref{lm2} above
instead). In particular (\ref{eq:50}) equals $2\pi\sigma_B$ as
defined. This proves (\ref{eq:41}). To prove (\ref{eq:42}), we let $b\le a/2$
and note that
by (\ref{eq:27}, \ref{eq:27a}, \ref{eq:46})
\begin{eqnarray*}
\lefteqn{\Vert (Q_a-P) (P-P^2+Q_a-Q_a^2)\Vert_1}\hspace{3cm}\\
&&\le \Vert (Q_a-P)(1-F_b)(P-P^2+Q_a-Q_a^2)\Vert_1 + 2\Vert
(Q_a-P) F_b\Vert_1\\
&&\le C(1+a) \e^{-\kappa b} + C_N (1+a)^{-N}\;.
\end{eqnarray*}
Upon choosing e.g. $b=a^{1/2}$, this tends to 0 as
$a\to\infty$. \hfill $\square$\\

As a preparation to the proof of Lemma~\ref{lm5} we have: 

\begin{lem}\label{lm7} Eq.~(\ref{eq:1}) is well-defined and independent of 
$\chi$ and $g$ as stated in Theorem~\ref{thm}. In particular, 
\begin{equation}
\sigma_E=-\lim_{a\to\infty} \tr(g'(H) \i[H,\chi_a(x)])\;,
\label{eq:8}
\end{equation}
where $\chi_a(x)=\chi(x/a)$.
\end{lem}

\noindent
{\bf Proof.} Eq.~(\ref{eq:1}) is well-defined by (\ref{eq:A.8a}). 
By taking differences of switch functions, independence amounts to
\[
\hbox{(i)}\quad \tr \bigl(g'(H) \i [H,X(x)]\bigr)=0\;,\qquad
\hbox{(ii)}\quad \tr\bigl( G'(H) \i [H, \chi (x)]\bigr)=0\;,
\]
where $X,G\in C_0^{\infty}({\mathbb R})$ with $\supp G\subset\Delta$. These 
statements are verified as follows:

i) Since $g'(H) X(x)\in {\mathcal J}_1$ by (\ref{eq:A.6}, \ref{eq:A.7}) we 
have $\tr\bigl( g'(H) [H,X]\bigr)=\tr \bigl( g'(H)
 HX\bigr) - \tr \bigl(g'(H) XH\bigr) = 0$ by cyclicity. This already proves 
(\ref{eq:8}).

ii) $\bigl([G(H),\chi]\bigr) (\vec{r}_1,\vec{r}_2)= G(H)
  (\vec{r}_1,\vec{r}_2) \bigl( \chi (\vec{r}_2) - \chi (\vec{r}_1)\bigr)$.
By (\ref{eq:A.6}, \ref{eq:A.7}), $[G(H),\chi]\in {\mathcal J}_1$ and
hence
\begin{equation}
\tr [ G(H), \chi (x)]=0\;.
\label{eq:9}
\end{equation}
We then pick $\widetilde G\in C_0^{\infty}$ with $\supp\widetilde G\subset\Delta$
and $\widetilde G G=G$. Then (\ref{eq:9}) may also be written, using cyclicity
and (\ref{eq:A.8}) as
\[
\tr [\widetilde G G, \chi]=
\tr\bigl([G, \chi]\widetilde G\bigr)+
\tr\bigl([\widetilde G, \chi] G\bigr)
=\tr\bigl([H,\chi](G'\widetilde G+\widetilde G' G)\bigr)
=\tr\bigl([H,\chi]G'\bigr)\;.
\]
\rightline{$\square$}\\
\goodbreak

\noindent
{\bf Proof} of Lemma~\ref{lm5}. Let $A=Q_a-P$. Then, by (\ref{eq:A.4}) and
(\ref{eq:49}), 
\[
|A(\vec{r}_1,\vec{r}_2)| = \big| g(H)(\vec{r}_1,\vec{r}_2)
\bigl(\widehat{U}_a(\vec{r}_1) - \widehat{U}_a(\vec{r}_2)\bigr)\big| 
\le C_N \bigl( 1+|\vec{r}_1-\vec{r}_2|\bigr)^{-N} \
\frac{|\vec{r}_1-\vec{r}_2|}{a+|\vec{r}_1|}\;, 
\]
so that by (\ref{eq:A.7}) $\Vert A^3\Vert_1=\Vert A\Vert_3^3 \le C
a^{-1}$. This proves (\ref{eq:43}). \par
For $b\le a/2$ we have
\[
F_b(y) \widehat{U}_a(\vec{r}) = F_b (y) \e^{2\pi\i \chi_a (x)}\;,
\]
where $\chi_a(x)=\chi(x/a)$ is a switch function. We then have, using
(\ref{eq:27}, \ref{eq:27a}),
\begin{eqnarray}
\lefteqn{\frac 3 2\tr \{ Q_a-P, P-P^2+Q_a-Q_a^2\}}\label{eq:28}\\
&& =3\tr F_b (Q_a-P) F_b (P-P^2+Q_a-Q_a^2) + O ((1+a)\e^{-\kappa b})
\nonumber\\
&&=3\tr F_b \bigl( P(2\pi) - P(0)\bigr) F_b \bigl(
P(0)-P(0)^2+P(2\pi)-P(2\pi)^2 \bigr) + O ((1+a)\e^{-\kappa b})\;,
\nonumber
\end{eqnarray}
where $P(\varphi) = \e^{\i\varphi\chi_a(x)} g(H) \e^{-\i\varphi\chi_a(x)}$.
We now apply the fundamental theorem of calculus to
\begin{equation}
P(2\pi)-P(0)
= \int_0^{2\pi} d\varphi \, \frac{d}{d\varphi} P(\varphi)
= - \int_0^{2\pi} d\varphi\, \e^{\i\varphi\chi_a} \i [g(H),\chi_a]
\e^{-\i\varphi\chi_a}\;. 
\label{eq:29}
\end{equation}
We remark that in (\ref{eq:27}, \ref{eq:27a}, \ref{eq:25},
\ref{eq:26}) one can, by the same proof, replace $Q_a-P$ by
$\i[g(H),\chi_a]:$
\begin{eqnarray}
&\Vert \i [g(H),\chi_a] F_b\Vert_2 \le C (b/a)^{1/2}\;,&
\label{eq:30}\\
&\Vert \i[g(H),\chi_a] (1-F_b)(g(H)-g(H)^2)\Vert_1\le C(1+a)\e^{-\kappa b}\;.&
\label{eq:31}
\end{eqnarray}
Thus
\[
\sup_{0\le \varphi, \varphi' \le 2\pi}  \Vert \bigl(
P(\varphi')-P(\varphi)\bigr) F_b\Vert_2 \le C(b/a)^{1/2}\;, 
\]
so that by writing
\begin{eqnarray*}
\lefteqn{\bigl( P(\varphi')-P(\varphi')^2\bigr) - \bigl(
P(\varphi)-P(\varphi)^2\bigr)  =}\hspace{5cm}\\
&&\bigl( P(\varphi')-P(\varphi)\bigr)
\bigl( 1-P(\varphi')\bigr) - P(\varphi) \bigl( P(\varphi')-P(\varphi)\bigr)
\end{eqnarray*}
we infer
\[
\sup_{0\le\varphi,\varphi'\le 2\pi} \Vert F_b \bigl[\bigl(
P(\varphi')-P(\varphi')^2\bigr) - \bigl(
P(\varphi)-P(\varphi)^2\bigr)\bigr] F_b\Vert_2 \le C (b/a)^{1/2}\;.
\]
Using this with $\varphi'=0,2\pi$, (\ref{eq:29}, \ref{eq:30}) and the
Cauchy-Schwarz inequality we find that (\ref{eq:28}) equals, up to errors
$O (b/a)+O((1+a)\e^{-\kappa b})$,
\begin{eqnarray*}
\lefteqn{-6 \int_0^{2\pi} d\varphi\,\tr \bigl( F_b \i [g(H),\chi_a] F_b
\bigl( g(H)-g(H)^2\bigr)\bigr) }
\hspace{2.5cm} 
\\
&&= -6 \cdot 2\pi \,\tr \bigl(\i[g(H),\chi_a] \bigl(
g(H)-g(H)^2\bigr)\bigr) \\
&&=-2\pi\cdot 6\, \tr \bigl( \i[H,\chi_a] g' (H) \bigl(
g(H)-g(H)^2\bigr)\bigr)=
- 2\pi\,\tr \bigl( \i[H,\chi_a] \widetilde{g}'(H)\bigr)\;,
\end{eqnarray*}
where $F_b$ has been dropped using (\ref{eq:31}) and (\ref{eq:A.8}) been used.
We remark that $\widetilde{g}=3g^2-2g^3$ is also a switch function. We 
finally pick e.g. $b=a^{1/2}$ so that the error mentioned above vanishes as 
$a\to\infty$. Thus (\ref{eq:22}) follows from Lemma~\ref{lm7}.
\hfill$\square$\\

\noindent
{\bf Proof} of Lemma.~\ref{lm1}. This is a variant of the argument leading
to (\ref{eq:12}) in the case of projections. Let, in the general case,
$A,B$ be as in (\ref{eq:16}, \ref{eq:21}). Then, by (\ref{eq:13}),
\begin{eqnarray*}
f_t (A)-A^3&=&(1-R_t)(\hbox{r.h.s. of (\ref{eq:19})}) \\
&=& \frac 1 4 \bigl( (1-R_t)[B,[B,A]]-R_t \{B,\{A,B\}\}\bigr)\\
&&-\frac 1 2 R_t \{ A,1-A^2-B^2\}\\
&&+\frac 14 \{ B,\{A,B\}\} - \frac 14 \{A, 1-A^2-B^2\}\\
&&+\frac 34 \{ A, 1-A^2-B^2\}\\
&\equiv& L_1 +L_2+L_3+L_4\qquad \hbox{(linewise)},
\end{eqnarray*}
where
\begin{equation}
1-R_t = tA^2 (1+tA^2)^{-1}\;,
\label{eq:32}
\end{equation}
resp. $R_t=(1+tA^2)^{-1}$. Note that since $A=A^*$ 
\begin{eqnarray}
&&\slim_{t\to\infty} t^{1/2} A R_t = 0\;,
\label{eq:33}\\
&&\slim_{t\to\infty} R_t = \Pi\;, 
\label{eq:34}
\end{eqnarray}
where $\Pi$ is the projection onto the null space of $A$. 

1) We claim $\lim_{t\to\infty} \tr L_1=0$. To this end we
  consider the first term in the corresponding bracket first:
\begin{equation}
(1-R_t)[B,[B,A]] = [ B,(1-R_t)[B,A]] + [B,R_t][B,A]\;.
\label{eq:36}
\end{equation}
Since $A\in {\mathcal J}_3$ and $1-R_t \in {\mathcal J}_{3/2}$ we have
$(1-R_t) [B,A] \in {\mathcal J}_1$ by the H\"older inequality. Thus
$\tr [B, (1-R_t)[B, A]]=0$. The last term in (\ref{eq:36}) is by
(\ref{eq:20}) 
\begin{eqnarray*}
[B,R_t][B,A]& =& t R_t [A^2,B] R_t \cdot [B,A] \\
&=& t R_t [A,\{A,B\}] R_t [B,A]\\
&=& t R_t A \{A,B\} R_t (-2AB +\{ A,B\})- t R_t \{A,B\} A R_t (2BA-\{ A,B\})\\
&=& - 2t R_t A\{A,B\} R_t AB
-2t R_t \{ A,B\} A R_t B A\\
&&+ t R_t \{ A, \{A,B\}\} R_t \{ A,B\}\;. 
\end{eqnarray*}
All terms are trace class since $\{A,B\}$ is
by (\ref{eq:17}, \ref{eq:6d}) with
$p(\lambda)=2(\lambda-\lambda^2)$. We recall that 
\begin{eqnarray}
X_n\displaystyle\mathop{\to}^s 0,\  Y \in {\mathcal J}_1 \ \Rightarrow \
\Vert X_n Y\Vert_1 \to 0\;, \nonumber\\
X_n^* \displaystyle\mathop{\to}^s 0, \  Y \in {\mathcal J}_1 \
\Rightarrow \ \Vert Y X_n\Vert_1 \to 0\;.
\label{eq:37}
\end{eqnarray}
Thus the first two terms on the r.h.s. do not contribute to the trace
as $t\to\infty$ by (\ref{eq:33}) (use cyclicity for the
second). Similarly, in the last term
\[
t R_t (A^2B+2ABA+BA^2) R_t \{ A,B\}\;,
\]
the middle term thereof does not. Using cyclicity of the trace on the
remaining ones, as well as (\ref{eq:32}), we find for $t\to\infty$
\begin{eqnarray*}
\tr (1-R_t) [B, [B,A]]
&=& \tr B R_t \{A, B\} (1-R_t) + \tr B (1-R_t) \{ A,B\}
R_t + o(1)\\
& =& \tr B R_t \{ A,B\} + \tr B \{A,B\} R_t + o(1) \\
& =& \tr R_t \{ B, \{A,B\}\} + o(1)\;,
\end{eqnarray*}
where we used $R_t \{A,B\}R_t
\displaystyle\mathop{\longrightarrow}_{t\to\infty} \Pi \{A,B\} \Pi=0$
in trace norm, a consequence of (\ref{eq:34}, \ref{eq:37}). The traces
of the two terms in $L_1$ thus compensate one another in the limit
$t\to\infty$. 

2) We note that $\{A, 1-A^2-B^2\} \in {\mathcal J}_1$ by
(\ref{eq:16}, \ref{eq:18}, \ref{eq:6c}). Again by (\ref{eq:34}) we have 
\[
- 2\lim_{t\to\infty} \tr L_2 = \tr \Pi \{ A,
1-A^2-B^2\}
= \tr\Pi \{ A,1-A^2-B^2\} \Pi = 0
\]
since $\Pi=\Pi^2$.

3) $L_3$ equals the second line of the r.h.s. of (\ref{eq:14}),
as seen from (\ref{eq:19}). Hence $\tr L_3=0$ follows from
(\ref{eq:38}) for $p(\lambda)=(1-2\lambda) (\lambda-\lambda^2)$. 

We can now summarize:
\[
\lim_{t\to\infty} \tr f_t (A) = \tr A^3 + \frac 3 4 
\tr \{ A, 1-A^2-B^2\}\;,
\]
which is (\ref{eq:39}).\hfill$\square$\\

As a final remark, we note that $\lim_{t\to\infty} \tr f_t
(UPU^*-P)$, if existent, is invariant under trace class perturbations
of $U$. This follows from (\ref{eq:A.3}). Similarly, as a
possible replacement for Lemma~\ref{lm4}, one has, without making recourse to Lemma~\ref{lm1},
\[
\lim_{a\to\infty} \tr f_t \bigl(\widecheck{U}_a g(H) \widecheck{U}_a^* -
g(H)\bigr) = 2\pi\sigma_B
\]
uniformly in $t\ge 1$. This follows from the proof of Lemma~\ref{lm4}
together with (\ref{eq:12}) and (\ref{eq:A.2}).

\begin{appendix} 
\numberwithin{equation}{section}

\section{Appendix}

\begin{alem}\label{lmA1} Let $X=X^*, Y=Y^*$ and $t\ge 0$. For $X\in {\mathcal
  J}_3$,
\begin{equation}
\Vert f_t (X)\Vert_1 \le (1+t)\Vert X\Vert_3^3\;.
\label{eq:A.1}
\end{equation}
If $X-Y\in{\mathcal J}_1$, then
\begin{equation}
\Vert f_t (X)-f_t(Y)\Vert_1 \le 3(1+t^{-1}) \Vert X-Y\Vert_1
\label{eq:A.2}
\end{equation}
and
\begin{equation}
\lim_{t\to\infty} \tr \bigl( f_t (X)-f_t(Y)\bigr) = \tr(X-Y)\;. 
\label{eq:A.3}
\end{equation}
\end{alem}

\noindent
{\bf Proof.} Eq. (\ref{eq:A.1}) is evident from (\ref{eq:5}). From
\[
f_t (\lambda) = (1+t^{-1}) \bigl[ \lambda -
\frac{\lambda}{1+t\lambda^2}\bigr] 
\]
and from
\[
X(1+tY^2)-(1+tX^2) Y=X-Y- tX(X-Y)Y
\]
we find
\[
f_t(X)-f_t(Y)=(1+t^{-1})\bigl[ X-Y-(1+tX^2)^{-1} (X-Y-tX(X-Y)Y)
(1+tY^2)^{-1}\bigr]\;. 
\]
Using 
$\Vert (1+tX^2)^{-1} \Vert \le 1,\, \Vert t^{1/2} X (1+tX^2)^{-1}\Vert\le 1$
we obtain (\ref{eq:A.2}). Using furthermore
\begin{eqnarray*}
&&\slim_{t\to\infty} t^{1/2} X (1+tX^2)^{-1} = 0\;,\\
&&\slim_{t\to\infty} (1+tX^2)^{-1} = \Pi_X\;, 
\end{eqnarray*}
where $\Pi_X$ is the projection onto the null space of $X$, together
with (\ref{eq:37}), we obtain
\[
f_t (X)-f_t(Y) \displaystyle\mathop{\longrightarrow}_{t\to\infty}
X-Y-\Pi_X(X-Y)\Pi_Y=X-Y 
\]
in trace norm. \hfill $\square$

\begin{alem}\label{lmA3} For $1\le p < \infty$,
\begin{equation}
\Vert T\Vert_p \le \sum_{\vec{s}} \Bigl(\sum_{\vec{r} \in {\mathbb Z}
  \times {\mathbb N}} | T (\vec{r}+\vec{s},\vec{r})|^p\Bigr)^{1/p}\;.
\label{eq:A.7}
\end{equation}
\end{alem}
\noindent
{\bf Proof.} The case $p=3$ is Eq. (4.11) in \cite{AG}, and the proof given
there applies to $1\le p < \infty$. \newline\phantom{ } \hfill $\square$

\begin{alem}\label{lmA2} i) Let $g\in C^\infty ({\mathbb R})$ with $\supp g'$ 
compact. Then, for any $N$,
\begin{equation}
|g(H) (\vec{r}_1,\vec{r}_2)| \le C_N (1+|\vec{r}_1-\vec{r}_2|)^{-N}\;. 
\label{eq:A.4}
\end{equation}
ii) If furthermore $\supp g'\subset \Delta $, then, for
some $\kappa > 0$,
\begin{equation}
\big|\bigl( Jg(H)J^*-g(H_B)\bigr) (\vec{r}_1,\vec{r}_2)\big| \le C_N
(1+|\vec{r}_1-\vec{r}_2|)^{-N} \e^{-\kappa\min (|y_1|,|y_2|)}\;,
\label{eq:A.5}
\end{equation}
unless both $y_1,y_2<0$.

\noindent
iii) If $G\in C_0^{\infty} ({\mathbb R})$ with $\supp G \subset
\Delta$, then 
\begin{equation}
\big| G(H) (\vec{r}_1,\vec{r}_2)\big| \le C_N
(1+|\vec{r}_1-\vec{r}_2|)^{-N} \e^{-\kappa (y_1+y_2)} .
\label{eq:A.6}
\end{equation}
In particular, $\e^{\kappa y} G(H)$ is a bounded operator. 
\end{alem}

\begin{alem}\label{lmA4} Let $\chi', g',G\in C_0^{\infty}$ with 
$\supp G\subset\Delta$. Then 
\begin{equation}
[H,\chi(x)]G(H)\;,\quad[g(H),\chi(x)]G(H) \in{\mathcal J}_1
\label{eq:A.8a}
\end{equation}
and
\begin{equation}
\tr([g(H),\chi(x)]G(H))=\tr([H,\chi(x)]g'(H)G(H))\;.
\label{eq:A.8}
\end{equation}
\end{alem}

\noindent
In \cite{D}, Chapter 2, or \cite{HS}, Lemma~B.1 the Helffer-Sj\"ostrand formula
\begin{equation}
g(H)=\frac{1}{2\pi}\int_{{\mathbb R}^2}\partial_{\bar z}
\tilde g(z)(H-z)^{-1}dxdy\;,\qquad (z=x+\i y)\;,
\label{eq:hs}
\end{equation}
is proven in the sense of a norm convergent integral for $H$ a self-adjoint
operator on a Hilbert space $\mathcal H$ and, say, $g\in C_0^{\infty}$, where 
$\partial_{\bar z}=\partial_x+\i\partial_y$ and $\tilde g$ is a quasi-analytic
extension of $g$. For definiteness, let 
\[
\tilde g(z)=\sum_{k=0}^Ng^{(k)}(x)\frac{(\i y)^k}{k!}\chi(y)\;,
\]
with $N\ge 1$, and hence
\begin{equation}
\partial_{\bar z}\tilde g(z)=
g^{(N+1)}(x)(\i y)^N\chi(y)+
\i\sum_{k=0}^Ng^{(k)}(x)\frac{(\i y)^k}{k!}\chi'(y)\;,
\label{eq:dg}
\end{equation}
where $\chi\in C_0^{\infty}$ is even and equals $1$ in a neighborhood
$(-\delta,\delta)$ of
$y=0$. In Lemma~\ref{lmA2} one is mainly interested in functions with 
$\supp g'$, but not $\supp g$, compact. The difference is of little importance, since, if $H$ were bounded above or below, one could trade the one for the other by adding a constant to $g$ and changing it outside of the spectrum. As we however do not want to resort to this assumption, we maintain that (\ref{eq:hs}) still holds in the 
strong sense.\\

\noindent
{\bf Proof} of Lemma~\ref{lmA2}. We claim that
\begin{equation}
g(H)\psi=\frac{1}{2\pi}\int dx
\Bigl(\int dy\partial_{\bar z}\tilde g(z)(H-z)^{-1}\psi\Bigr)
\label{eq:shs}
\end{equation}
for all $\psi\in{\mathcal H}$ and $g'\in C_0^{\infty}$. By the functional 
calculus it suffices to show that, if $\psi$ is dropped and $H$ replaced 
by $a\in{\mathbb R}$, the r.h.s. is (a) well-defined as an improper Riemann
integral, and (b) agrees with $g(a)$. Indeed, all of 
$\partial_{\bar z}\tilde g$, except for the $k=0$ term in (\ref{eq:dg}),
has compact support $K\subset{\mathbb R}^2$, and 
\begin{equation}
|\partial_{\bar z}\tilde g(z)-\i g(x)\chi'(y)|\le C|y|^N\;,
\label{eq:qa}
\end{equation}
so that the analysis of \cite{D,HS} still applies, except for the 
contribution 
from $\i g(x)\chi'(y)$. The latter equals, using that $\chi'$ is odd,
\begin{eqnarray*}
\lefteqn{\frac{\i }{2\pi}\int dxg(x)
\Bigl(\int_0^\infty dy\chi'(y)[(a-x-\i y)^{-1}-(a-x+\i y)^{-1}]\Bigr)}
\hskip 3cm\\ \hskip 3cm&&
=-\frac{1}{\pi}\int dx g(x)\Bigl(\int_0^\infty dy\,y\chi'(y)[(a-x)^2+y^2]^{-1}
\Bigr)\;,
\end{eqnarray*}
which is absolutely convergent. This proves (a); part (b) follows as, 
e.g., in \cite{D,HS}.

Let $R(\vec{r}_1,\vec{r}_2;z)=(H-z)^{-1}(\vec{r}_1,\vec{r}_2)$ be the Green
function. We shall use the Combes-Thomas \cite{CT} estimates
\begin{eqnarray}
|R(\vec{r}_1,\vec{r}_2;x+\i y)|&\le& 
\frac{2}{|y|}\e^{-\mu|\vec{r}_1-\vec{r}_2|}\;,
\label{eq:ct1}\\
\int |R(\vec{r}_1,\vec{r}_2;x+\i y)-R(\vec{r}_1,\vec{r}_2;x-\i y)|dx &\le& 
12\sqrt{2}\pi\e^{-\mu|\vec{r}_1-\vec{r}_2|}\;,
\label{eq:ct2}
\end{eqnarray}
which hold true provided
\begin{equation}
\sup_{\vec{r}_0\in{\mathbb Z}\times {\mathbb N}}\sum_{\vec{r}}
|H(\vec{r}_0,\vec{r}_0+\vec{r})|(\e^{\mu|\vec{r}|}-1)
\le |y|/2\;.
\label{eq:ctc}
\end{equation}
They have been proven in this form in \cite{AG}, Appendix D,
Eqs. (D.3, D.4, D.11). Since 
$(\e^{\mu|\vec{r}|}-1)\le (\mu/\mu_0)(\e^{\mu_0|\vec{r}|}-1)$ for
$0\le\mu\le\mu_0$ we may take, by (\ref{eq:srh}), $\mu=c|y|$ for
$y\in\supp\chi$, where $c>0$ is some small constant. 

i) The contribution to (\ref{eq:A.4}) from the $k=0$ term in (\ref{eq:dg}) is,
through (\ref{eq:shs}),
\[
\frac{\i}{2\pi}\int dx g(x) \Bigl(\int_0^\infty dy\chi'(y)
\bigl(R(\vec{r}_1,\vec{r}_2;x+\i y)-R(\vec{r}_1,\vec{r}_2;x-\i y)\bigr)\Bigr)\;,
\]
and is bounded in modulus by 
\[
6\sqrt{2}\|g\|_\infty\int_0^\infty dy|\chi'(y)|\e^{-c|y| |\vec{r}_1-\vec{r}_2|}
\le C\e^{-c\delta|\vec{r}_1-\vec{r}_2|}\;,
\]
by using (\ref{eq:ct2}). The remaining contribution is bounded using 
(\ref{eq:ct1}, \ref{eq:qa}) as
\[
C\int_K dxdy|\chi'(y)||y|^N\frac{2}{|y|}\e^{-c|y||\vec{r}_1-\vec{r}_2|}
\le C_N(1+c|\vec{r}_1-\vec{r}_2|)^{-N}\;,
\]
since $K$ is compact. 

ii) It suffices to establish a bound of the form $C\e^{-2\kappa |y_1|}$ if
$y_2\ge 0$ for the l.h.s. of (\ref{eq:A.5}). In fact by applying that estimate
to $\bar g$ we can interchange $y_1$ and $y_2$ in the bound, and hence replace
it by $C\e^{-2\kappa\min (|y_1|,|y_2|)}$ for $y_1,\,y_2$ as specified in the
lemma. Moreover, we can also bound 
(\ref{eq:A.5}) by a constant times $(1+|\vec{r}_1-\vec{r}_2|)^{-2N}$ in virtue
of (\ref{eq:A.4}), which applies to $H_B$ as well. Then (\ref{eq:A.5}) follows
since $\min(a,b)\le (ab)^{1/2}$ for $a,b>0$. 

For $y_2\ge 0$ the matrix element (\ref{eq:A.5}) is 
$(Jg(H)-g(H_B)J)(\vec{r}_1,\vec{r}_2)$. We use the resolvent identity
$J(H-z)^{-1}-(H_B-z)^{-1}J=-(H_B-z)^{-1}E(H-z)^{-1}$
in (\ref{eq:shs}) and
distinguish as before between the contribution, $I$, to (\ref{eq:A.5}) from 
$\i g(x)\chi'(y)$, and the rest, $II$. Using again that $\chi'$ is odd and
\begin{eqnarray*}
\lefteqn{(H_B-z)^{-1}E(H-z)^{-1}-(H_B-\bar z)^{-1}E(H-\bar z)^{-1}=}&&\\
&&[(H_B-z)^{-1}-(H_B-\bar z)^{-1}]E(H-z)^{-1}+
(H_B-\bar z)^{-1}E[(H-z)^{-1}-(H-\bar z)^{-1}]
\end{eqnarray*}
we have 
\begin{eqnarray*}
\lefteqn{I=-\frac{\i}{2\pi}\int dx g(x) \Bigl(\int_0^\infty dy\chi'(y)
\sum_{\tl{\vec{r}\in{\mathbb Z}^2}{ \vec{r}^{\,\prime}\in{\mathbb Z}\times{\mathbb N}}}
\Delta R_B(\vec{r}_1,\vec{r};x+\i y)
E(\vec{r},\vec{r}^{\,\prime})R(\vec{r}^{\,\prime},\vec{r}_2;x+\i y)}&&\\
&&\hskip 6cm
+R_B(\vec{r}_1,\vec{r};x-\i y)E(\vec{r},\vec{r}^{\,\prime})
\Delta R(\vec{r}^{\,\prime},\vec{r}_2;x+\i y)
\Bigr)\;,
\end{eqnarray*}
where 
$\Delta R(\vec{r}_1,\vec{r}_2;z)= 
R(\vec{r}_1,\vec{r}_2;z)-R(\vec{r}_1,\vec{r}_2;\bar z)$. We use (\ref{eq:ct1})
for $R,\, R_B$ and (\ref{eq:ct2}) for $\Delta R,\,\Delta R_B$, and bound 
$\e^{-c\delta|\vec{r}^{\,\prime}-\vec{r}_2|}$ by $1$. The result is
\begin{eqnarray*}
|I|&\le&
\sum_{\tl{\vec{r}\in{\mathbb Z}^2}{\vec{r}^{\,\prime}\in{\mathbb Z}\times{\mathbb N}}}
|E(\vec{r},\vec{r}^{\,\prime})| |F_I(\vec{r}_1,\vec{r},\vec{r}^{\,\prime},\vec{r}_2)|\;,\\
|F_I|&\le&\frac{\|g\|_\infty}{2\pi}\Bigl( \int_0^\infty dy|\chi'(y)|\Bigr)
12\sqrt{2}\pi\cdot\frac{2}{\delta}\e^{-c\delta|\vec{r}_1-\vec{r}|}\cdot 2\;,
\end{eqnarray*}
so that by (\ref{eq:Eest})
$|I|\le C\sum_{\vec{r}\in{\mathbb Z}^2}
\e^{-\mu_0|y|}\e^{-c\delta|\vec{r}_1-\vec{r}|}$.
We may at this point assume $\mu_0< c\delta$ and use 
$|y|\ge|y_1|-|\vec{r}_1-\vec{r}|$, so that
\begin{equation}
|I|\le C\e^{-\mu_0|y_1|}\sum_{\vec{r}\in{\mathbb Z}^2}
\e^{-(c\delta-\mu_0)|\vec{r}_1-\vec{r}|}\le C\e^{-\mu_0|y_1|}\;.
\label{eq:I}
\end{equation}

Before turning to $II$ we note that $|y|$ in (\ref{eq:ct1}, \ref{eq:ctc}) can
be replaced with $\dist(x+\i y,\sigma(H))$. This follows by inspection of the
proof, Eqs. (D.8-D.10) in \cite{AG}. By the spectral condition (\ref{eq:0})
and the assumption of (ii) we
have $\dist(z,\sigma(H_B))\ge d$ for some $d>0$ and all 
$z\in\supp \partial_{\bar z}\tilde g$. Therefore,
\[
II=-\frac{1}{2\pi}\int_K dxdy(\partial_{\bar z}\tilde g(z)-\i g(x)\chi'(y))
\sum_{\tl{\vec{r}\in{\mathbb Z}^2}{ \vec{r}^{\,\prime}\in{\mathbb
Z}\times{\mathbb N}}}
R_B(\vec{r}_1,\vec{r};x+\i y)E(\vec{r},\vec{r}^{\,\prime})
R(\vec{r}^{\,\prime},\vec{r}_2;x+\i y)
\]
can be estimated as 
\begin{eqnarray*}
|II|&\le&
\sum_{\tl{\vec{r}\in{\mathbb Z}^2}{\vec{r}^{\,\prime}\in{\mathbb Z}
\times{\mathbb N}}}
|E(\vec{r},\vec{r}^{\,\prime})| 
|F_{II}(\vec{r}_1,\vec{r},\vec{r}^{\,\prime},\vec{r}_2)|\;,\\
|F_{II}|&\le& C\int_K dxdy|y|^N\frac{2}{d}\e^{-cd|\vec{r}_1-\vec{r}|}
\cdot\frac{2}{|y|}\le C\e^{-cd|\vec{r}_1-\vec{r}|}\;.
\end{eqnarray*}
We conclude as in (\ref{eq:I}).

iii) In this case $G(H_B)=0$, and (\ref{eq:A.6}) follows from (\ref{eq:A.5}).
The final remark follows e.g. from Holmgren's bound \cite{Fri}: 
$\|A\|\le\max(\sup_{\vec{r}_1}\sum_{\vec{r}_2}|A(\vec{r}_1,\vec{r}_2)|, 
\sup_{\vec{r}_2}\sum_{\vec{r}_1}|A(\vec{r}_1,\vec{r}_2)|)$.\hfill $\square$\\

\noindent
{\bf Proof} of Lemma~\ref{lmA4}. By 
$\|[H,\chi]G(H)\|_1\le \|[H,\chi]\e^{-\kappa y}\|_1\|\e^{\kappa y}G(H)\|$ and 
(\ref{eq:A.6}) we are left to show that $T=[H,\chi]\e^{-\kappa y}$ is trace
class. Its kernel is 
\[
T(\vec{r}_1,\vec{r}_2)=H(\vec{r}_1,\vec{r}_2)(\chi(x_2)-\chi(x_1))
\e^{-\kappa y_2}\;.
\]
Since
\[
|\chi(x_2)-\chi(x_1)|\le C\frac{|x_2-x_1|(1+|x_2-x_1|)}{1+x_2^2}
\le C\frac{\e^{\mu_0|x_2-x_1|}-1}{1+x_2^2}
\]
we have by (\ref{eq:srh})
\[
\sum_{\vec{s}} | T (\vec{r}+\vec{s},\vec{r})|\le 
C\frac{\e^{-\kappa y}}{1+x^2}\;,
\]
which is summable w.r.t. $\vec{r}=(x,y)\in {\mathbb Z} \times {\mathbb N}$. 
The first part of (\ref{eq:A.8a}) thus follows by (\ref{eq:A.7}) with $p=1$. 
The same proof with $H$ replaced by $g(H)$, except that (\ref{eq:A.4}) is 
used instead of (\ref{eq:srh}), implies the second part of (\ref{eq:A.8a}).

Eq. (\ref{eq:shs}) implies, see \cite{HS}, Eqs. (B.10, B. 14), 
\begin{eqnarray}
[g(H),\chi]&=&-\frac{1}{2\pi}\int dx
\Bigl(\int dy\partial_{\bar z}\tilde g(z)(H-z)^{-1}[H,\chi](H-z)^{-1}\Bigr)
\;,
\label{eq:shs1}\\
g'(H)&=&-\frac{1}{2\pi}\int dx
\Bigl(\int dy\partial_{\bar z}\tilde g(z)(H-z)^{-2}\Bigr)\;,
\label{eq:shs2}
\end{eqnarray}
where the integrals are again meant in the strong sense. For the two sides of 
(\ref{eq:A.8}) we may write
\begin{eqnarray*}
\tr([g(H),\chi(x)]G(H))&=&
\tr(E_\Delta(H)[g(H),\chi(x)]G(H)E_\Delta(H))\;,\\
\tr([H,\chi(x)]g'(H)G(H))&=&
\tr(E_\Delta(H)[H,\chi(x)]g'(H)G(H)E_\Delta(H))\;.
\end{eqnarray*}
We now multiply (\ref{eq:shs2}) from the left by 
$[H,\chi]$, and both (\ref{eq:shs1}, \ref{eq:shs2}) by $E_\Delta(H)$
from the left and by $G(H)E_\Delta(H)$ from the right. The integrals then become
absolutely convergent in trace class norm. This follows from (\ref{eq:qa}) and
from
\begin{eqnarray*}
\|E_\Delta(H-z)^{-1}[H,\chi](H-z)^{-1}GE_\Delta\|_1&\le&
\|[H,\chi]G\|_1\|(H-z)^{-1}E_\Delta\|^2\;,\\
\|E_\Delta[H,\chi](H-z)^{-2}GE_\Delta\|_1&\le&
\|[H,\chi]G\|_1\|(H-z)^{-2}E_\Delta\|\;,
\end{eqnarray*}
since 
$\|(H-x-\i y)^{-p}E_\Delta\|\le C|x|^{-p}$ for large $x$. The traces can
thus be carried inside the integral representations, where they are seen to be
equal by cyclicity.\hfill $\square$\\

After completion of this work we learned from A. Klein that Lemma \ref{lmA2}(i) appeared in \cite{GK}.

\end{appendix}

\end{document}